\documentclass{emulateapj}
\usepackage{graphics}

\slugcomment{Accepted to \textit{The Astrophysical Journal}}

\begin{document}


\shortauthors{Temim et al.}

\shorttitle{``G327.1-1.1 with Chandra, XMM''}

\title{Chandra and XMM Observations of the Composite Supernova Remnant G327.1-1.1}

\author{TEA TEMIM\altaffilmark{1,2}, PATRICK SLANE\altaffilmark{1}, B. M. GAENSLER\altaffilmark{3}, JOHN P. HUGHES\altaffilmark{4}, ERIC VAN DER SWALUW\altaffilmark{5}}

\altaffiltext{1}{Harvard-Smithsonian
Center for Astrophysics, CfA} 
\altaffiltext{2}{Department of Astronomy, School of Physics and
Astronomy, University of Minnesota}
\altaffiltext{3}{Institute of Astronomy, School of Physics, The University of
Sydney, NSW 2006, Australia} 
\altaffiltext{4}{Rutgers University}
\altaffiltext{5}{Royal Netherlands Meteorological Institute (KNMI), The Netherlands}

\begin{abstract}

We present new X-ray imaging and spectroscopy of a
composite supernova remnant G327.1-1.1 using the Chandra and XMM-Newton
X-ray observatories. G327.1-1.1 has an unusual morphology consisting
of a symmetric radio shell and an off center nonthermal component
that indicates the presence of a pulsar wind nebula (PWN). Radio
observations show a narrow finger of emission extending from the PWN
structure towards the northwest. X-ray studies with ASCA,
ROSAT, and BeppoSAX revealed elongated extended emission and a
compact source at the tip of the finger that may be coincident with
the actual pulsar. The high resolution Chandra observations provide
new insight into the structure of the inner region of the remnant.
The images show a compact source embedded in a cometary
structure, from which a trail of X-ray emission extends in the
southeast direction. The Chandra images also reveal two prong-like
structures that appear to originate from the vicinity of the compact
source and extend into a large bubble that is oriented in the north-west direction, opposite from the bright radio PWN. The emission from the entire radio shell is detected in
the XMM data and can be characterized by a thermal plasma model with
a temperature of $\sim$ 0.3 keV, which we use to estimate the physical properties of the remnant.
The peculiar morphology of G327.1-1.1 may
be explained by the emission from a moving pulsar and a relic PWN
that has been disrupted by the reverse shock.

\end{abstract}


\section{INTRODUCTION} \label{intro}

Composite supernova remnants (SNRs) are those for which we see
distinct evidence of the two fundamental components that
characterize the aftermath of massive star collapse. The blast wave
from the explosion sweeps up ISM material and heats it to X-ray
emitting temperatures while accelerating electrons that produce
radio synchrotron radiation in the compressed magnetic field. The highly
magnetic, rapidly rotating neutron star (NS) that is left behind
produces a particle wind which sustains an extended broadband synchrotron
nebula of magnetic flux and relativistic particles. X-ray
observations provide the thermal characteristics of the shell which
allow us to constrain the explosion energy, age, and the surrounding
ISM density. Simultaneously, the spectral and spatial properties of
a pulsar wind nebula (PWN) allow us to infer the properties of the
nebular pressure and magnetic field, and provide constraints on the
central pulsar created in the explosion.

G327.1-1.1 is a composite SNR that was originally discovered as non-thermal radio source by \citet{cla73,cla75}. It contains a bright central PWN whose structure is complex in both radio and X-ray bands. The radio morphology \citep{whi96} shows a faint shell, 17 arcmin
in diameter, surrounding a bright non-thermal PWN component that is
presumably powered by a yet-to-be-discovered pulsar. The bright part of the PWN is
located off-center with respect to the SNR shell. A
distinct finger-like structure protrudes from the PWN in the
northeast direction, possibly suggesting a picture in which a
fast-moving pulsar is moving through the SNR, leaving the PWN in its
wake. This type of PWN morphology indicated that the reverse shock of the SNR has disrupted the PWN \citep{blo01,swa04}. The X-ray emission from G327.1-1.1 was first detected by \citet{lam81}, who noticed an offset between the peak radio and X-ray emission. The composite nature of G327.1-1.1 was further confirmed by recent X-ray studies, which provide evidence of a compact X-ray
source that is the likely counterpart for the pulsar powering the
PWN emission \citep{sla98,sun99}.

There has also been evidence for thermal X-ray emission in
G327.1-1.1 in previous studies, but the properties of this component
are not well determined. ROSAT PSPC observations by \cite{sew96}
reveal X-ray emission concentrated at the position of the PWN with
some very faint emission extending outward toward the SNR shell.
Using a simple blast-wave interpretation, they conclude that the
remnant age is 7000 yr, although this value relies strongly on the
inferred temperature of the outer emission component which is not
well determined from the ROSAT data. ASCA observations also provide
weak evidence for thermal emission \citep{sun99} with kT $\sim$ 0.4 keV.
Using a Sedov model, their results imply a remnant age of 11000 yr,
a preshock density of $n_0=0.1\:cm^{-3}$, and a swept-up mass of 50
M$_{\sun}$. BepoSAX observations indicate a temperature of about 0.2
keV, however \citep{boc03}, which leads to an age of 29000 yr,
$n_0 \sim 0.4\:cm^{-3}$, and a swept-up mass of 800 M$_{\sun}$.

In this paper, we present \textit{Chandra} and \textit{XMM} X-Ray
imaging and spectroscopy of G327.1-1.1, along with the Molonglo
Observatory Synthesis Telescope (MOST) 843 MHz observations
\citep{whi96} for comparison. Observations and data reduction are
described in Section \ref{obsv} and imaging and spectral analysis in
Section \ref{analysis}. Section \ref{disc} discusses the nature of
the thermal and non-thermal emission, remnant and PWN properties and
morphology, and the evolutionary phase of G327.1-1.1. The
conclusions are summarized in Section \ref{concl}.

\section{OBSERVATIONS AND DATA REDUCTION} \label{obsv}

G327.1-1.1 was observed with the Advanced CCD Imaging Spectrometer, ACIS-I,  
on board the \textit{Chandra} X-ray observatory on 2001, July
15, under the observation ID 1955 and a total exposure time of 50
kiloseconds. The standard data reduction and cleaning were performed
in Ciao Version 3.4. The XMM-Newton observations were carried out on
2004, Feb 07 with the MOS1, MOS2, and PN cameras for a total
exposure time of 100 ks, under the observation ID 0203820101. The
MOS cameras were operated in the Full Frame Mode with a ``medium''
filter setting, and the PN camera was operated in the Small Window
Mode. The standard reduction of the data was performed using the
XMM-SAS software, version 7.1.0, and resulted in a final exposure
time of 85 ks for each of the MOS detectors, and 82 ks for the PN detector.


\begin{figure}
\epsscale{1.0} \plotone{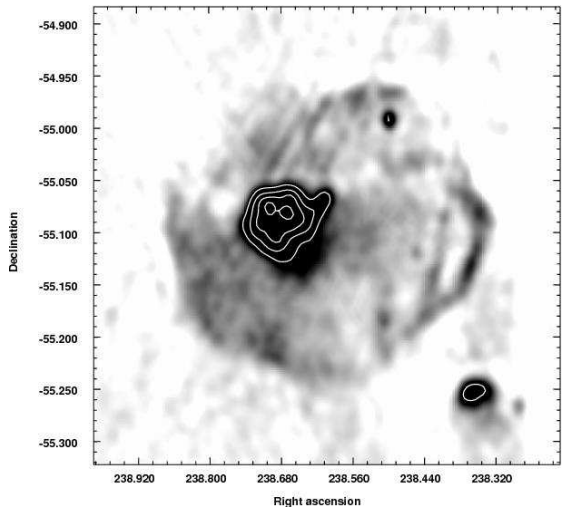} \caption{\label{rad}MOST 843 MHz radio
image of G327.1-1.1 with a resolution of 43\arcsec and the PWN region contours overlayed , with levels of 0.05, 0.09, 0.13, and 0.21 Jy/beam \citet{whi96}. The
image shows a faint shell, 17 arcmin in diameter, surrounding a
bright non-thermal component, from which a narrow finger of emission
extends to the northwest.}
\end{figure}


\subsection{Imaging}

In order to analyze the X-ray morphology of G327.1-1.1, images in
various energy bands were created from the \textit{Chandra} data. 
Point sources in the field were
subtracted from the cleaned event file using the Ciao task
\textit{dmfilth}. The missing pixel values for each point source
region were replaced by a Poisson distribution whose mean was
determined from the pixel values of the background region
surrounding each point source. X-ray images and the corresponding
exposure maps, were created with a binning factor of 8 and convolved
with a Gaussian function with sigma = 3 pixels. Finally, the binned
and smoothed images were divided by the exposure maps in order to
correct for the differences in the spectral response and effective
area across the field.


\begin{figure*}
\epsscale{1.0} \plottwo{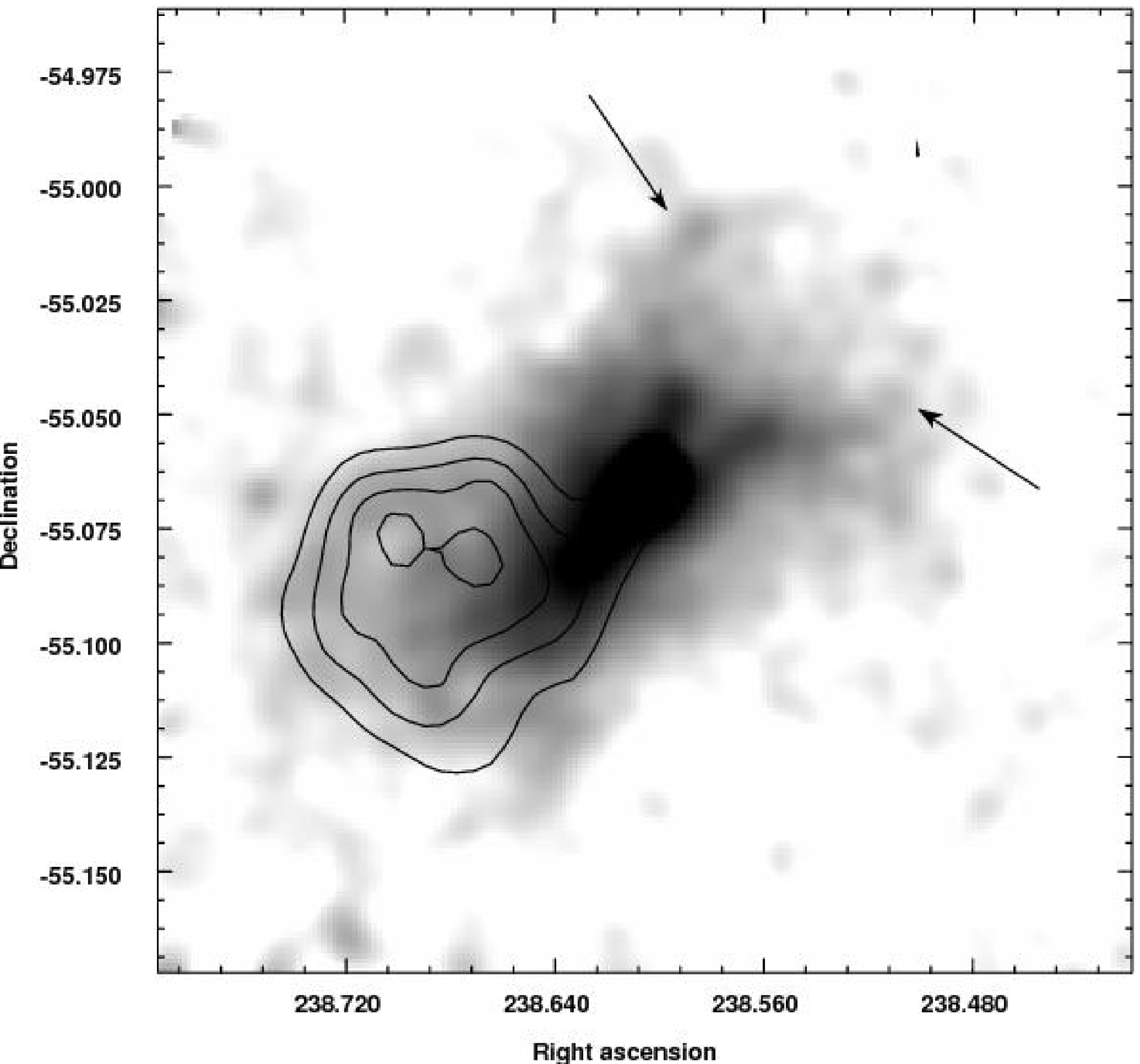}{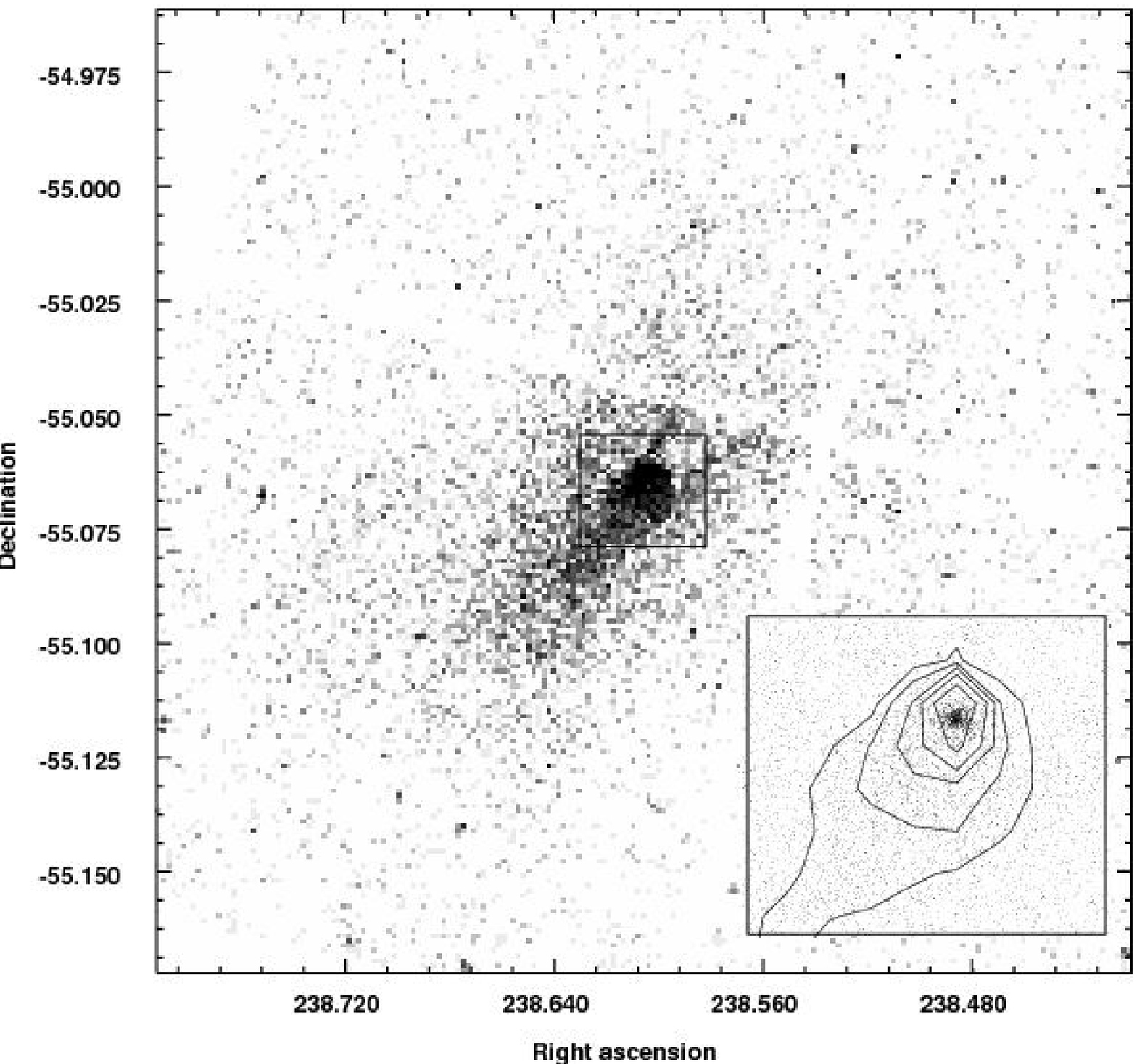}
\caption{\label{chandra}The left panel is the exposure corrected and smoothed
\textit{Chandra} ACIS image of G327.1-1.1 in the 0.5-10.0 keV band, with the contours of the radio image in Figure \ref{rad} overlayed. The image reveals a compact source at the
tip of the radio finger, from which a trail of X-ray emission extends
towards the southeast. A pair of bright prong-like structures originates from the
vicinity of the compact source and extends into a large bubble-like structure, indicated by the arrows.The left panel shows the same image, binned by a factor of 8, but unsmoothed. The compact source region in the black box, approximately 1\farcm5 by 1\farcm5 in size, is zoomed in at the bottom right corner of the image. The image of the zoomed in region is unbinned and unsmoothed. The contours are the binned and smoothed contours from the same data.}
\end{figure*}


Images in multiple X-ray energy bands were also created from the
cleaned \textit{XMM-Newton} MOS1 and MOS2 event files. Events from
the regions that correspond to bright point sources in the field
were removed from the data and were not refilled in this case. The
images and the exposure maps in the corresponding energy bands were
created using the SAS (version 7.1.0) software, with the spatial
binning size set to 5\arcsec. Since the background in the XMM
images includes a non-vignetted component from the internal
background and X-ray fluorescence, dividing by the exposure map
tends to leave a hollow ring in the images. In order to remove the
non-vignetted background component, we used the Filter Wheel Closed
(FWC) data provided by the \textit{XMM-Newton} EPIC Background
Working Group at the University of Birmingham (http://www.sr.bham.ac.uk/xmm3/BGproducts.html), which are dominated by the internal instrumental
background. FWC images of the internal background were created and
subtracted from our data, after correcting for the difference in
exposure times. Before subtracting the background and dividing by
the exposure maps, all the images were smoothed by a Gaussian
function with a width of 3 pixels (15\arcsec). Finally, the MOS1 and MOS2 images
were combined using the SAS task \textit{emosaic} to produce one
final MOS image.


\begin{figure}
\epsscale{1.0}
\plotone{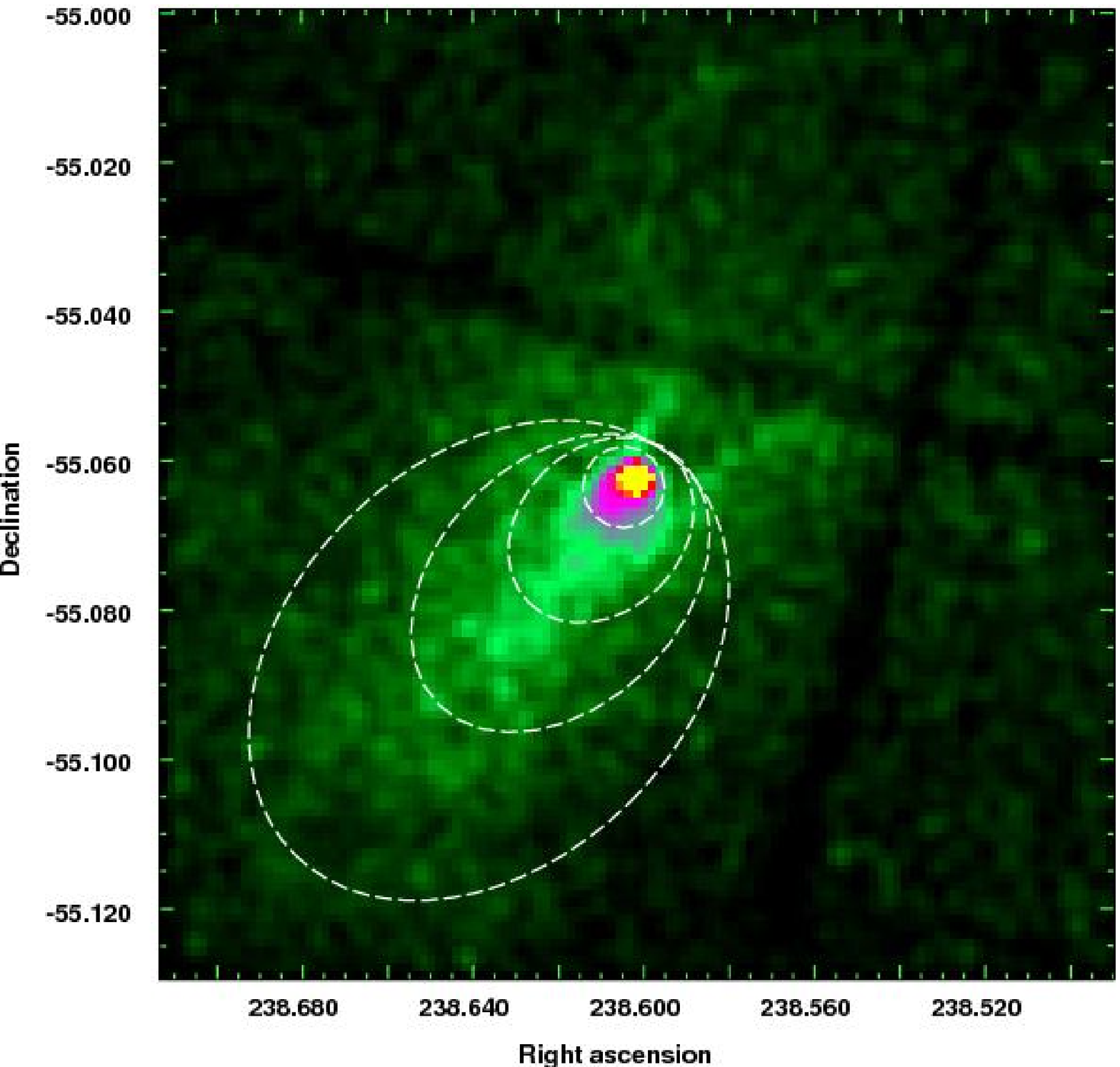}\caption{\label{bow}\textit{Chandra} ACIS
image of the inner region of G327.1-1.1 in the 0.5-10.0 keV energy band. The image was smoothed with
a 2 pixel Gaussian function and shows a compact source,
embedded in a cometary structure. The white dashed regions are
apertures from which spectra were extracted for the \textit{Chandra}-ACIS data (see Section \ref{spectroscopy}).}
\end{figure}




\begin{deluxetable*}{lcc}
\tablecolumns{3} \tablewidth{0pc} \tablecaption{\label{xmm}XMM
SPECTRAL FITTING RESULTS} \tablehead{ \colhead{PARAMETER} &
\colhead{PWN} & \colhead{Shell}} \startdata
$N_H$($10^{22}cm^{-2}$) & 1.91 $\pm$ 0.07 & (1.91) \\
Photon Index & 2.11 $\pm$ 0.03 & \nodata \\
kT (keV) & 0.29 $\pm$ 0.01 & 0.30 $\pm$ 0.01 \\
Residual background temperature (keV) & (3.3) & 3.3 $\pm$ 0.4 \\
$F_X$(total observed)($erg/cm^{2}/s$) & 7.3E-12 & 4.0E-12 \\
$F_X$(unabsorbed non-thermal)($erg/cm^{2}/s$) & 1.7E-11 & \nodata \\
$F_X$(unabsorbed thermal)($erg/cm^{2}/s$) & 9.8E-11 & 2.1E-10 \\
$F_X$(unabsorbed background)($erg/cm^{2}/s$) & 1.3E-12 & 5.1E-12 \\
Reduced $\chi^2$ Statistic & 1.08 & 1.26 \\
\enddata
\tablecomments{The listed uncertainties are 1.6 sigma (90 \% confidence) statistical uncertainties only. The apertures used in the spectral extractions are shown in Figure \ref{aps}, and the spectra and fits are shown in Figure \ref{spec}. The values in parantheses were held fixed. The fluxes were calculated in the 0.3-9.0 keV band and represent the average of the MOS1 and MOS2 flux values.}
\end{deluxetable*}




\begin{figure}
\epsscale{1.0} \plotone{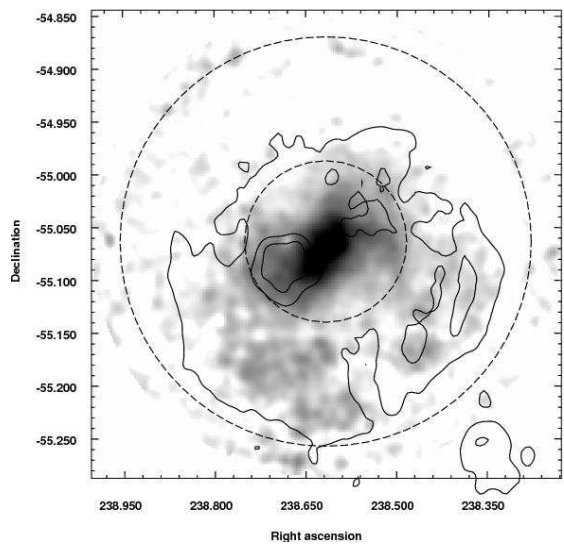} \caption{\label{aps}XMM-Newton
combined MOS1 and MOS2 image of G327.1-1.1, in the 1.0-2.0 energy
band, overlayed with the MOST radio contours. X-ray emission from
almost the entire shell is detected in the XMM data and coincides
well with the radio contours. The dashed circles in black are the spectral extraction regions for
the XMM data (see Section \ref{spectroscopy}).}
\end{figure}


\subsection{Spectroscopy}\label{spectroscopy}

\textit{XMM-Newton} MOS spectra were extracted from the cleaned and
point source subtracted event files using the SAS (version 7.1.0)
\textit{especget} task, which generates spectra and the
corresponding effective area and spectral response files. The
spectra were extracted from a circular aperture with a radius of 5\farcm2, 
centered on the PWN region of G327.1-1.1, and an annulus
with an inner radius of 5\farcm2 and and an outer radius of 11.6
\arcmin that mostly includes emission from the thermal shell. These
regions are shown in Figure \ref{aps}. Since there was no suitable
region available for background extraction due to the spatial extent
of the source, we used the combination of blank sky background and
FWC data to estimate the background. The blank sky background event
files were produced by the XMM-Newton EPIC Background Working Group
at the University of Birmingham, using observations for many sky positions \citep{car07}. The XMM
Newton \textit{Skycast} script was used to project the blank sky and
FWC event files into sky coordinates and match it to our dataset.
Spectra were then extracted from the same regions as the source
spectra and corrected to the same exposure times. The blank sky
background data is a combination of the sky background and the
internal instrumental background, which can vary as a function of
time. In order to determine the possible difference in the internal
background levels between our observations and the background
datasets, which were taken on different dates, we measured and
compared the high energy (10-12 keV) fluxes at the positions of the
detector that were out of the field of view. The blank sky high
energy flux was found to be lower by 5\% and 3\% for the MOS1 and
MOS2 observations, respectively. The spectra extracted from the FWC
data were then multiplied by these factors and added to the blank
sky spectra in order to account for the different levels in the
internal background and produce a final background spectrum. The
final backgrounds were then subtracted from the source spectra and
grouped to include a minimum of 100 and 50 counts per bin for the
PWN and the shell emission spectrum, respectively.

The \textit{Chandra} ACIS spectra and the corresponding effective
area and spectral response files were produced using the
\textit{specextract} script in Ciao version 3.4. The spectra were
extracted from elliptical apertures shown in Figure \ref{bow}, with
each larger aperture excluding the counts from the inner
apertures. Since the emission from the remnant covered the majority
of the ACIS detector, corresponding background spectra from the same
regions were extracted from the ACIS blank sky background files (http://cxc.harvard.edu/ciao/threads/acisbackground/). The
blank sky data are produced from observations at various sky
positions with high Galactic latitude and have been reprojected to
the same sky coordinates as the source event file using the aspect
solution for our observation. The background was
subtracted and the resulting source spectra were grouped to include
30 counts in each bin for the two smaller apertures and 50 counts in each bin for the two larger apertures. Since G327.1-1.1 is located near the Galactic plane, the blank sky spectra most
likely underestimate the background at lower energies, which may
result in a residual Galactic background component.

\section{ANALYSIS}\label{analysis}

\subsection{X-ray Morphology}

The Chandra ACIS raw and smoothed and exposure-corrected X-ray images of
G327.1-1.1 in the 0.3-10.0 keV band are shown in Figure \ref{chandra}.
The MOST radio image is shown in Figure \ref{rad} for comparison.
The radio image shows a symmetric radio shell, 17 arcminutes in
diameter, and a bright, off-center non-thermal component from which
a narrow finger of emission extends in the northwest direction. The
\textit{Chandra} images show a compact source at the tip of this
finger and more diffuse X-ray emission elongated in the direction
along the finger. The emission from the compact source does not have
a profile of a point source, but is slightly extended and may
consist of a point source imbedded in a more extended region. Figure
\ref{bow} shows the detailed structure of this region. The image is binned by a factor of 8, smoothed with a 2 pixel Gaussian, and has not been exposure corrected. The compact source is embedded
in a cometary structure from which an X-ray trail extends in
the southeast direction. A pair of prong-like structures appears to originate
from the vicinity of the compact source and extends out to the
northwest. The most unusual feature in the X-ray images is a
bubble-like structure that extends out from the prongs. The bubble
is approximately 3 arcminutes in diameter and can be seen in both
Chandra and XMM images. The bubble is indicated by the black arrows in Figure \ref{chandra}a and is also evident in Figure \ref{bubble}, where
the unsmoothed Chandra image is shown in blue/green and the radio image in red. 

In the XMM MOS images, we detect the emission from nearly the entire radio
shell. The combined MOS1 and MOS2 image in the 1.0-2.0 keV band,
overlayed with the radio contours, is shown in Figure \ref{aps}. The
X-ray emission coincides well with the radio contours of the thermal
shell. As in the radio image, the non-thermal component in X-rays is
displaced from the geometric center of the X-ray shell.



\begin{deluxetable*}{lcccc}
\tablecolumns{5} \tablewidth{0pc} \tablecaption{\label{chandratab}CHANDRA ACIS SPECTRAL FITTING RESULTS}
\tablehead{
\colhead{PARAMETER} & \colhead{Region 1} & \colhead{Region 2} & \colhead{Region 3} & \colhead{Region 4}}
\startdata
Area (arcsec$^2$) & 1216 & 6124 & 15395 & 40105 \\
Photon Index & 1.65 $\pm$ 0.13 & 1.80 $\pm$ 0.13 & 2.03 $\pm$ 0.12 & 2.14 $\pm$ 0.14 \\
$F_X$(total observed)($erg/cm^{2}/s$) & 7.8E-13 & 7.9E-13 & 9.1E-13 & 1.0E-12 \\
$F_X$(unabsorbed non-thermal)($erg/cm^{2}/s$) & 1.4E-12 & 1.6E-12 & 2.2E-12 & 2.7E-12 \\
$F_X$(unabsorbed thermal)($erg/cm^{2}/s$) & \nodata & 2.1E-12 & 4.0E-12 & 1.5E-11 \\
Reduced $\chi^2$ Statistic & 0.6 & 0.7 & 1.0 & 0.6 \\
\enddata
\tablecomments{The extraction regions are shown in Figure \ref{bow}. Listed uncertainties are 1.6 sigma (90 \% confidence) statistical uncertainties from the fit. The spectra and fits are shown in Figure \ref{spec2} and the fluxes were calculated in the 0.3-9.0 keV band.}
\end{deluxetable*}



\subsection{Spectroscopy}

\begin{figure*}
\epsscale{1.0} \plottwo{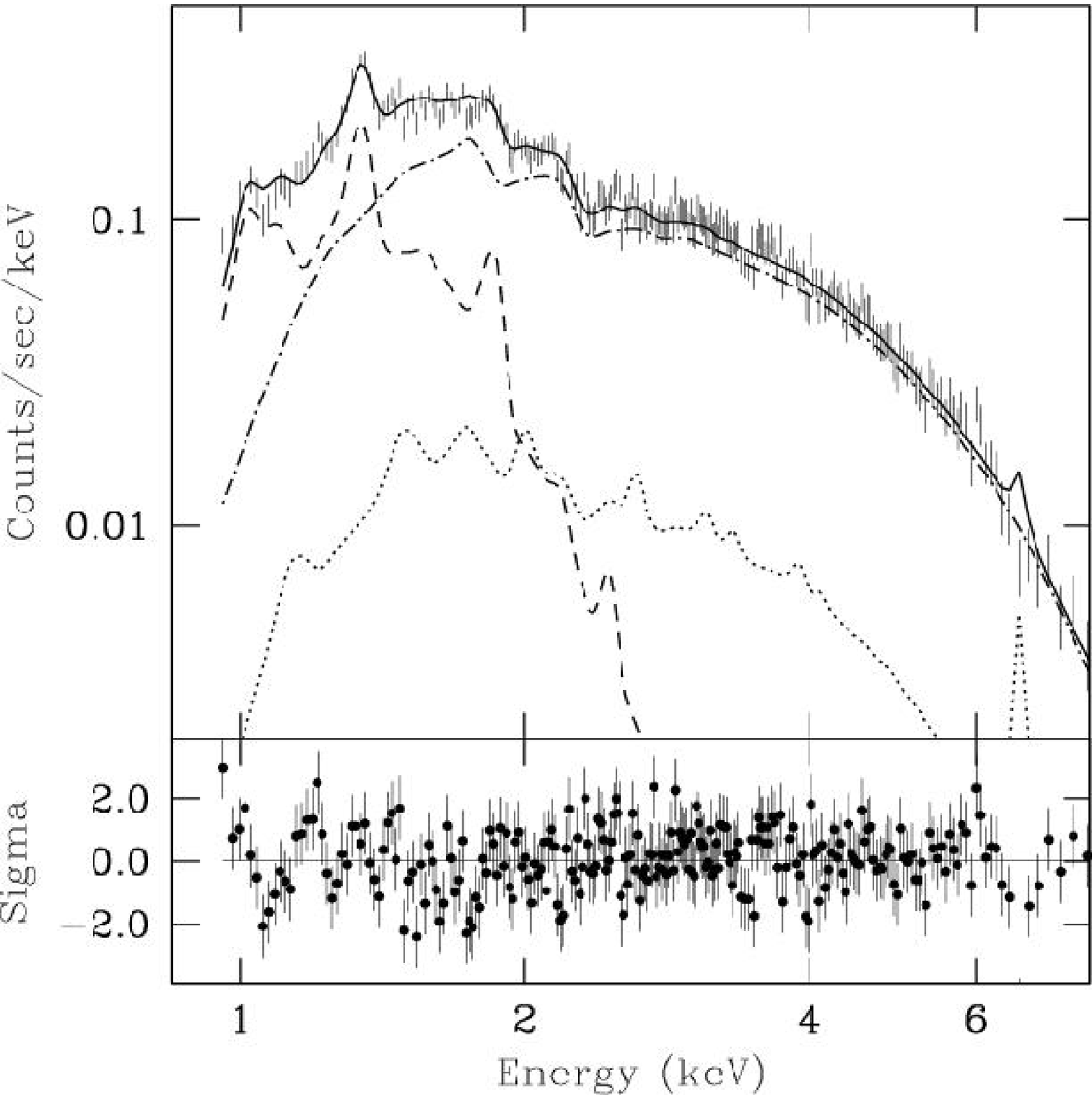}{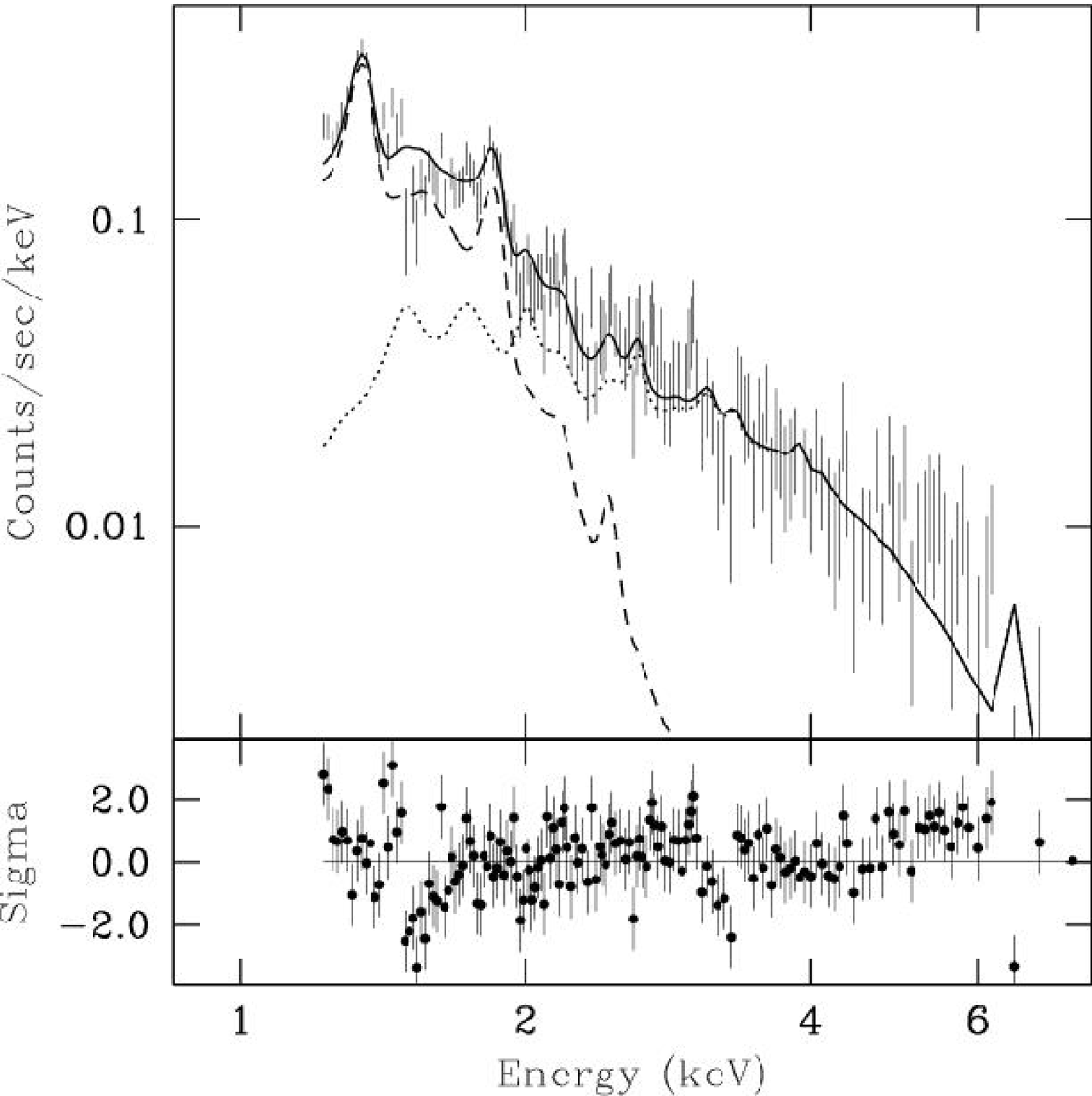}
\caption{\label{spec}Spectra extracted from the XMM MOS1 data of the PWN
region (left) and shell region (right). The solid lines represent
the two component best fits to the spectra with the individual
components shown as dotted and dashed curves. The PWN components are a
power-law (dot-dashed curve), 0.3 keV thermal model (dashed curve), and the 3.3 keV thermal model attributed to the residual Galactic background (dotted curve). The shell components are a 0.3 keV thermal model (dashed curve) and the same 3.3 keV thermal residual background (dotted curve). The extraction apertures are shown as dashed circles in Figure \ref{aps}. The PWN spectrum was extracted
from the inner circle and the shell spectrum from the annulus
defined by the two circles.  The best fit parameters are listed in
Table \ref{xmm}.}
\end{figure*}

In order to determine the general spectral properties of the X-ray
emission in G327.1-1.1, spectra were extracted from two different
regions in the XMM data, the inner PWN region and the outer shell. A
circular aperture 5\farcm2 in diameter, centered at 15:54:28.2, -55:03:49.25, was
used for the PWN region, and an annulus with the same center
coordinates and an inner and outer radii of 5\farcm2 and 11\farcm6
was used for the shell region (see Figure \ref{aps}). The spectra
were fit with the Ciao 3.4 Sherpa software and are shown in Figure
\ref{spec}. The spectra extracted from MOS1 and MOS2 data were fit
simultaneously. The absorption column density, photon indices of the
power law models, and temperatures of the thermal models were linked
for the MOS1 and MOS2 fits, but the normalization parameters of each
model were left to vary.


The spectrum extracted from the annular aperture and the best fit
models are shown in the right panel of Figure \ref{spec}. The
emission appears to be thermal in nature and we see evidence of
emission lines from NeX, MgXI, and SiXIII. Since we expect most of
the emission in the annular aperture to be thermal emission from the
shell, we attempted to fit the spectrum with a one component XSPEC Raymond
\& Smith thermal plasma model with fixed cosmic abundances. While a one component thermal model
can account for most of the emission at lower energies, an
additional component is needed in order to account for the excess
emission at energies above 2 keV. Due to the high uncertainties in
this region of the spectrum, this additional component can be fit
equally well with a power law model with a photon index of 2.28 $\pm$ 0.16, or a thermal model with a temperature of 3.3 $\pm$ 0.4 keV. While we cannot rule of the possibility that this residual high energy emission is non-thermal emission from the PWN particles that have diffused out to the outer remnant, in the fitting we assume that this emission is thermal emission from the background.  In the final fit,
we used a two component thermal model, which resulted in a 0.29
$\pm$ 0.01 keV temperature component that accounts for most of the
total observed emission, and a 3.3 $\pm$ 0.4 keV temperature component,
most likely due to residual Galactic background emission that was
not accounted for by the blank sky background spectra. The summary
of the best fit parameters for the shell region is listed in
Table \ref{xmm}. The listed uncertainties are 90 \% confidence, formal statistical uncertainties only.

While the spectrum of the inner PWN region of G327.1-1.1 is
dominated by non-thermal emission with a power-law spectrum, it is
obvious that an additional component is needed at energies below 2
keV. The spectrum was fit by an absorbed power law plus a thermal
model, where the thermal model was the Raymond \& Smith
thermal plasma model with fixed cosmic abundances. The 3.3 keV thermal background component, found from the XMM spectral fitting of the shell region, was also added to the fit to account for the residual Galactic background. The temperature and normalization of the background component, scaled by the difference in extraction areas, were held fixed in the fit. Due do its small contribution, the fit parameters stay roughly the same even if this component is excluded from the fit. The left panel of
Figure \ref{spec} shows the PWN spectrum and the best fit model. The power-law component is shown as the dot-dashed curve, the thermal component as the dashed curve, and the fixed thermal background as the dotted curve. The best fit power-law model has a photon index of 2.11 $\pm$ 0.03 and the thermal component has a temperature of
0.29 $\pm$ 0.01 keV. The parameters are summarized in Table \ref{xmm}.

\begin{figure*}
\epsscale{0.8} \plotone{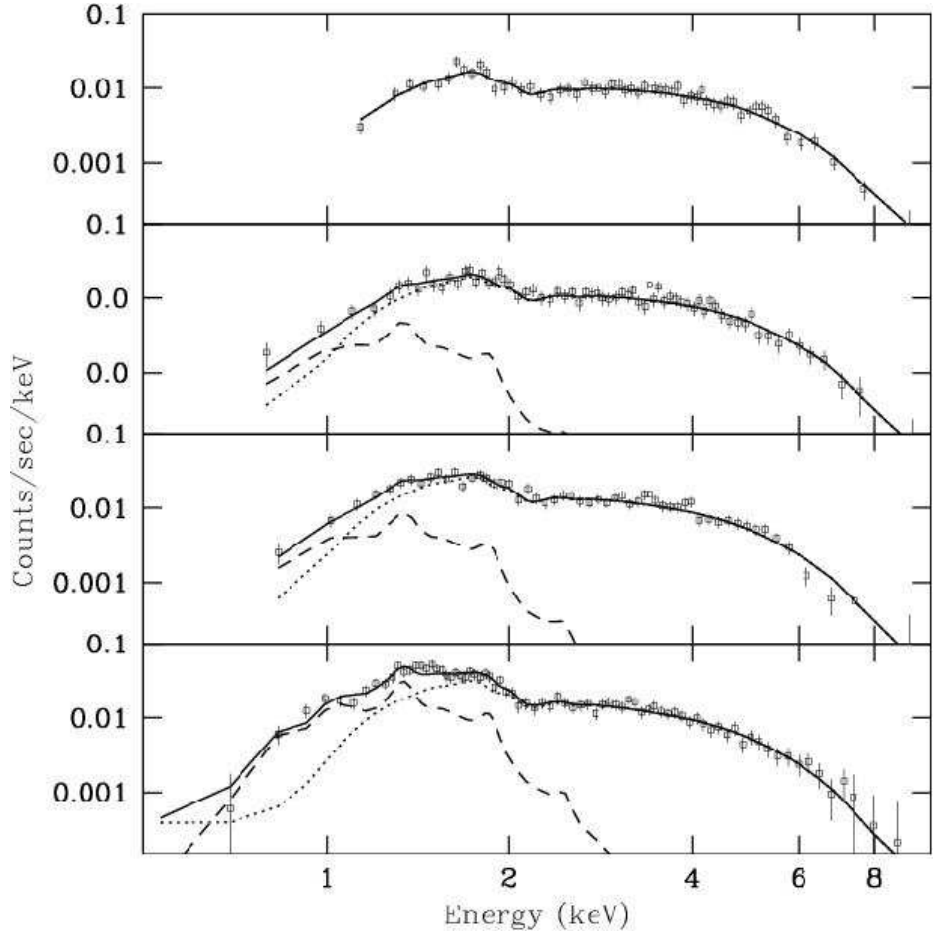} \caption{\label{spec2}Spectra
extracted from the \textit{Chandra} ACIS data using the elliptical
apertures shown in Figure \ref{bow}. Each of the
larger apertures excludes the counts from the inner ones and the
spectral plots (top to bottom) are ordered by the aperture size
(smallest to largest). The solid line represents the best fit two
component model, power-law and thermal. The individual components
are represented by dashed curves for the thermal component and
dotted curves for the power-law component. The best fit parameters
are listed in Table \ref{chandratab}.}
\end{figure*}

In order to determine how the spectrum of the non-thermal emission
varies with distance from the compact source, spectra were extracted
from the \textit{Chandra} ACIS event files using four circular
apertures shown in Figure \ref{bow}, where each aperture excludes
the counts from the enclosed apertures. For more details about
background subtraction, refer to Section \ref{spectroscopy}. Each of
the spectra were fit with an absorbed power law plus a Raymond \&
Smith thermal plasma model whose temperature was fixed to 0.29 keV,
as determined from the spectral fitting of the XMM-observed shell emission. The contribution from the residual 3.3 keV background component, found in the fitting of the XMM-observed shell emission, is neglegible compared to the non-thermal emission inside the ACIS apertures and was not included in the fit. The
value of N$_H$ was fixed to 1.91E22 $cm^{-2}$, the value found from the XMM data.
The spectra and best fit models are shown in Figure \ref{spec2} and
the best fit parameters are summarized in Table \ref{chandratab}. As
expected, the emission across the PWN is dominated by non-thermal
emission, but the spectrum extracted from the larger aperture
clearly requires an additional thermal component at energies below 2
keV. The best fit values for the photon indices show that the
spectrum is steepening with distance from the compact source. This
is consistent with the synchrotron burn-off of high energy electrons
at larger distances from the compact source.

\subsection{Timing Analysis}

The \textit{XMM} PN data in the small window mode (6 ms time resolution) was used to search for the pulsed signal from the compact source in G327.1-1.1. Events in the 0.5-10.0 keV and 2.0-7.0 keV bands were extracted from a circular aperture with a radius of 20\arcsec, centered on 15:54:24.5, -55:03:45.1 (see Section \ref{comps}). We used the $Z_{n}^{2}$ test \citep{buc83} with one harmonic to perform a timing analysis on a total of $\sim$6900 events between 0.5-10 keV and $\sim$4000 events between 2.0-7.0 keV, 50\% of which can be attributed to the compact source, and the rest to the local background. In Section \ref{comps}, we show that the spatial structure of the compact source in the \textit{Chandra} data is composed of a point source embedded in a more extended structure that accounts for the majority the emission. When we fit the extended component with a Gaussian, the point source, assumed to be the pulsar, accounts for less than 10\% of the total flux in the compact source.  This makes it undetectable above the local background, even with pulsed fraction of 100 \%. It may be possible that the point source contributes a larger fraction to the total flux, since the spatial profile of the underlying extended component is unknown. We searched for periodicity in the 1E-05 to 83 Hz frequency range, with a step size of 3E-06 Hz, a factor of 3 oversampling compared to 1/T, where T is the duration of the observation. We found no peaks with a probability below $10^{-3}$, representing a 3 sigma confidence level.

\section{DISCUSSION}\label{disc}

\subsection{Thermal Emission}\label{thermal}


XMM images of G327.1-1.1 reveal emission from the entire shell
structure observed in the radio. Figure \ref{aps} shows the XMM
image in the 1-2 keV band overlayed with the MOST radio contours,
which coincide extremely well with the X-ray emission. Spectral
fitting of the shell region indicates that the emission is thermal
and well described by a 0.3 keV thermal plasma. We used the best fit
temperature of the thermal shell to estimate the physical properties
of the remnant based on the \citet{sed59} simple blast wave model,
in which a supernova with the explosion energy of $E_0$ expands into
an ISM with a uniform density, $n_0$. We assume that the ISM is compressed into a
shell with four times the initial ambient density, and a thickness
of $1/12$ of the remnant radius, R. In our calculations, we also assume a
complete shell, a remnant radius of 8\farcm5, as measured from
the radio image, and equal ion and electron temperatures. The distance to G327.1-1.1 is not well determined and introduces the greatest uncertainty in the derived parameters.
We adopt a distance of 9.0 kpc, which was estimated by \citet{sun99}
based on the statistical relation between the hydrogen column
density in the X-ray band and E(B-V) \citep{ryt75}, and the relation
between E(B-V) and distance \citep{luc78}.

The following equations were used to calculate the remnant radius,
$R$, shock velocity, $v_s$, and the ISM density, $n_0$, where the last equation is based on the fitted intesnity of the X-ray spectrum and the assumed geometry. The equations are given as a function of the angular size of
the remnant in arcminutes, $R_{ang}$, the distance in kiloparsecs,
$D_{kpc}$, and the shell temperature, $T_{keV}$.

\begin{equation}
R(pc) = 0.291{R}_{ang}^{}{D}_{kpc}^{}
\label{R}
\end{equation}

\begin{equation}
{v}_{s}(km\:s^{-1})=928{T}_{keV}^{1/2}
\label{v}
\end{equation}

\begin{equation}
{n}_{0}(cm^{-3})=9.14{R}_{ang}^{-3/2}{D}_{kpc}^{-1/2}
\label{n}
\end{equation}

The \citet{sed59} model yields the following relations for the
remnant age, $t$, SN explosion energy, $E_0$, and the swept up mass, $M$.

\begin{equation}
t(yr)=124{R}_{ang}^{}{D}_{kpc}^{}{T}_{keV}^{-1/2}
\label{t}
\end{equation}

\begin{equation}
{E}_{0}(10^{51}erg)=2.87\times{10}^{-4}{R}_{ang}^{3/2}{D}_{kpc}^{5/2}{T}_{keV}^{}
\label{E}
\end{equation}

\begin{equation}
M({M}_{\odot}^{})=5.22\times10^{-3}{R}_{ang}^{3/2}{D}_{kpc}^{5/2}
\label{M}
\end{equation}

The derived values for G327.1-1.1 are listed in Table \ref{sedov}
and appear to be reasonable for a remnant in the
Sedov-Taylor stage. Since the model assumes a uniform ambient
density, and a single average thermal temperature, the derived
parameters have unquantifiable uncertainties due to uncertainties in
the distance and the likely inhomogeneities in the ambient ISM that
would cause variations in density and temperature across the shell.



\begin{deluxetable}{lc}
\tablecolumns{2} \tablewidth{0pc} \tablecaption{\label{sedov}INPUT AND DERIVED PROPERTIES FOR THE SEDOV MODEL} 
\tablehead{\colhead{Property} & \colhead{Value} \\} \startdata
D($kpc$) ~~~~~~~& 9.0 \\
T ($keV$) ~~~~~~~& 0.3 \\
R($pc$) ~~~~~~~& 22 \\
v($km/s$)~~~~~~~& 500 \\
t($yr$) ~~~~~~~& 18000 \\
n$_0$($cm^{-3}$) ~~~~~~~& 0.12 \\
M ($M_{\odot}^{}$) ~~~~~~~& 31 \\
E$_{51}$ ($10^{51}erg$) ~~~~~~~& 0.5 \\
\enddata
\tablecomments{A distance of 9.0 kpc was assumed in the
calculations.}
\end{deluxetable}



\subsection{Non-thermal Emission}\label{nonthermal}

The central part of G327.1-1.1 is dominated by a non-thermal
X-ray component whose total emission can be described by a power-law model
with a photon index of 2.08 $\pm$ 0.03, typical for a PWN. The X-ray
data provide evidence for the existence of a pulsar powering the
PWN. Based on the empirical relationship between the pulsar's
current spin-down energy loss rate and the luminosity of the
non-thermal emission \citep{sew88}, \citet{sun99} estimated a pulsar
period ($P$) of 62 ms, period derivative ($\dot{P}$) of
$8.9\times10^{-14}\:s\:s^{-1}$, and a surface magnetic field ($B_0$) of
$2.3\times10^{12}\:G$. In this section, we follow the discussion by \citet{sun99} and
derive the same pulsar and PWN properties for comparison. We assume
the same distance of 9.0 kpc and the remnant properties that we
derived in Section \ref{thermal}(Table \ref{sedov}).

The unabsorbed flux from the non-thermal component in the XMM data
is $8.0\times10^{-12}\:erg\:cm^{-2}\:s^{-1}$ in the 2-10 keV band. Based on
a more recent relationship between the pulsar's spin-down energy loss rate and the non-thermal luminosity, $logL_{X,(2-10keV)}=1.34log\dot{E}-15.3$ \citep{pos02}, and a distance of 9.0 kpc, this yields
a spin-down energy, $\dot{E}$, of $2.8\times10^{37}\:erg\:s^{-1}$. Using an SNR age of 18000 yr, we derive the following values; $P=35 ms$, $\dot{P}=3.1\times{10}^{-14}ss^{-1}$, and $B_0=1.0\times{10}^{12}G$. Due to the very large uncertainties associated with the distance and age of G327.1-1.1, the calculated values are only rough approximations and are consistent with previous estimates by \citet{sun99}.

\subsection{Evolution and Morphology}\label{evolution}


\begin{figure*}
\epsscale{0.9} \plotone{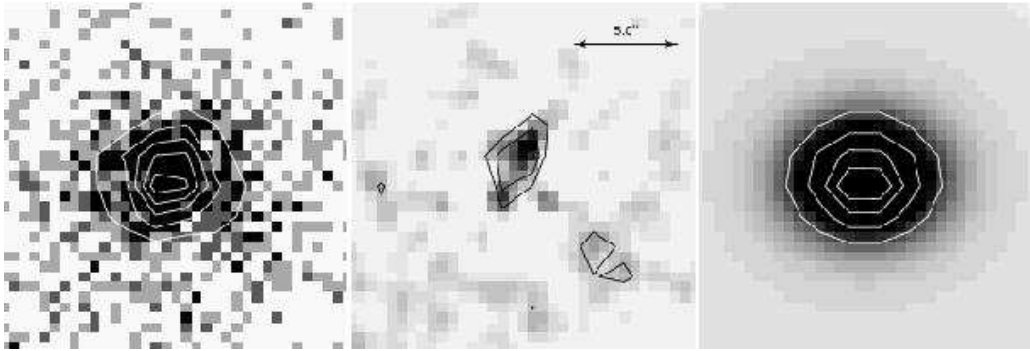} \caption{The left panel shows the unsmoothed \textit{Chandra} image of the compact source region. The right panel is the image of the best fit model to the compact source emission, consisting of a point source and a more extended Gaussian component. The middle panel is the residual image. All three panels are displayed on the same linear scale from 0-16 counts. \label{psimage}}
\end{figure*}


The evolution of a PWN inside a composite SNR can be divided into
three stages when the pulsar is not moving; the supersonic expansion stage, the reverse shock interaction stage, and the subsonic expansion stage. In the initial supersonic expansion stage, the PWN
is bounded by a strong shock as it expands into the surrounding SN
ejecta. The reverse shock eventually encounters the PWN surface and
crushes the nebula, causing it to reverberate. After the reverse
shock interaction stage, the PWN continues to expand subsonically
into the SN ejecta that has been heated by the reverse shock
\citep{swa01,blo01}. 

In the case where the pulsar is moving through
the SNR at a high velocity, as it appears to be the case for
G327.1-1.1, the evolution and morphology become much more complex. 
Initially, the PWN is carried along with the moving pulsar and
the reverse shock first interacts with the PWN surface that is
closest to the SNR shell. The PWN is swept away from the pulsar,
resulting in a relic PWN that is located at a position opposite of
the pulsar motion. After the passage of the reverse shock, the
moving pulsar continues to generate a PWN that is now expanding
subsonically into the reheated SN ejecta and is connected to the
old, relic PWN. An additional stage of PWN evolution becomes evident
when the pulsar's velocity becomes supersonic and the head of its
PWN deforms into a bow shock \citep{swa04}.

In the case of G327.1-1.1, we observe a bright radio PWN, which may
be interpreted as the older, relic PWN, left behind after the
passage of the reverse shock. \citet{swa04} gives an
expression for the timescale, $t_{col}$, for the reverse shock to
collide with the entire PWN surface,

\begin{equation}
t_{col}=1045E_{51}^{-1/2}\left(\frac{M_{ej}^{}}{M_{\odot}^{}}\right)^{5/6}n_{0}^{-1/3} yr,
\label{tcol}
\end{equation}

where $E_{51}^{}$ is the explosion energy in units of
${10}^{51}\:erg$, and $M_{ej}^{}$ is the ejected mass. Assuming an
ejected mass of $10M_{\odot}^{}$, and using the derived values from
Table \ref{sedov}, we find the timescale for the reverse shock
collision to be on the same order as the derived SNR age of 18000 yr. 
This suggests that the PWN is either still in the reverse shock interaction stage, or that the reverberations from the collision between the reverse shock and the PWN have died out and the expansion of the overall PWN structure has become subsonic. The displacement of the relic
PWN in G327.1-1.1 cannot be explained by the asymmetries due to the
pulsar motion alone. Its displacement is not aligned with the X-ray
trail or the radio finger, but is instead located further to the
east. In order to explain the displacement of the PWN from the center of the SNR shell, a combination of the pulsar motion and an asymmetric reverse shock may be required.

Chandra observations clearly show that the compact source is
surrounded by an extended, cometary structure, with an average
radius of approximately half an arcminute (see Figures \ref{chandra},\ref{bow}, and \ref{psplot}). According to hydrodynamic simulations of \citet{swa04},
there are two possible scenarios that may give rise to this type of
morphology. The pulsar may be moving at a supersonic velocity with
respect to the SNR, causing the PWN to deform into a true bow shock, or
the reverse shock has disrupted the PWN from the NW, giving rise to
the cometary morphology.

\subsubsection{Pulsar Velocity and Bow Shock Formation}\label{pulsar_vel}

The formation of the bow-shock is expected to occur at half the
crossing time, $t_{cr}$ (defined as  the age of the remnant when the pulsar reaches the
SNR shell), assuming that the remnant is in the Sedov-Taylor stage \citep{swa03}; 

\begin{equation}
{t}_{cr}\simeq1.4\times{10}^{4}E_{51}^{1/3}V_{1000}^{-5/3}n_{0}^{-1/3}yr,
\label{tcr}
\end{equation}

\noindent where $V_{1000}$ is the pulsar velocity in units of $1000\:km\:s^{-1}$. At this time, the pulsar
is positioned at a distance equal to two thirds of the blast wave
radius, $R$ \citep{swa04}.
Using the physical parameters for G327.1-1.1, we find that the
pulsar velocity would need to be $\sim\:770\:km\:s^{-1}$ in order for the
current age of the remnant to be larger than half the crossing time $t_{cr}$.
This is the pulsar velocity required for the bow-shock formation to
have already taken place in G327.1-1.1. The approximate displacement
of the compact source in G327.1-1.1 from the geometric center of its
radio shell is 7.5 parsecs. This is also the approximate length of
the radio finger, which is presumably the pulsar trail. Using this
projected distance and the derived age, we calculate the tangential
pulsar velocity to be on the order of $\sim\:400\:km\:s^{-1}$. This would require
a 30 degree angle between the pulsar velocity and the line of sight
to achieve a velocity of $770\:km\:s^{-1}$, required for bow-shock
formation. While the required pulsar velocity is somewhat high for a typical pulsar, it is not entirely unreasonable and cannot discount the possibility that the structure in Figure
\ref{bow} is the head of the PWN that has already been deformed into a bow shock.

\subsubsection{Spatial Modeling of the Compact Source}\label{comps}



\begin{deluxetable}{ll}
\tablecolumns{2} \tablewidth{0pc} \tablecaption{\label{compacttab}BEST-FIT PARAMETERS FOR THE \textit{CHANDRA} SPATIAL MODELING OF THE COMPACT SOURCE IN G327.1-1.1}
\tablehead{
\colhead{PARAMETER} & \colhead{Value}}
\startdata
\cutinhead{Delta Function}
Center coordinates & 15:54:24.5, -55:03:45.1 \\
Amplitude & 48 $\pm$ 19 $counts$ \\
Total flux & 48 $counts$ \\
\cutinhead{Gaussian}
Center coordinates & 15:54:24.4, -55:03:44.6 \\
FWHM & 5\farcs5$\pm$0\farcs5 \\
Ellipticity & 0.21$\pm$0.08 \\
Total flux & 749 $counts$ \\
\cutinhead{Background level}
Amplitude & 1.1 $counts\:arcsec^{-2}$ \\
\enddata
\end{deluxetable}



\begin{figure}
\epsscale{1.0} \plotone{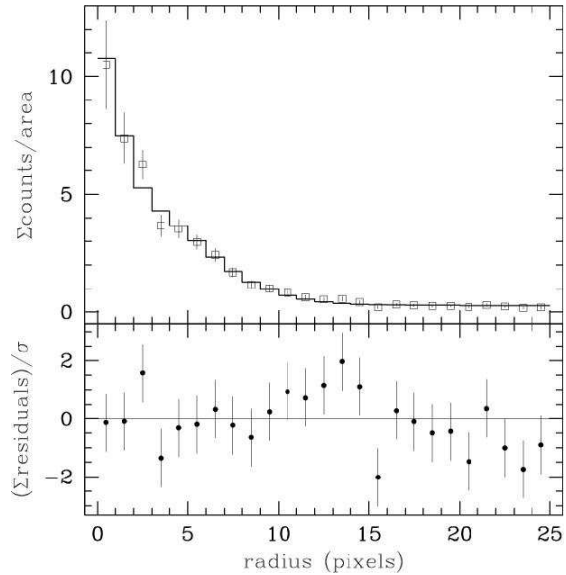}
\caption{\textit{Chandra} emission profile of the compact source region. The curve representing the best fit model composed of a \textit{Chandra} PSF and a more extended Gaussian component. The profile is centered at the coordinates 15:54:24.5, -55:03:45.1, and each pixel corresponds to a spatial size of 0\farcs5. \label{psplot}}
\end{figure}

\begin{figure}
\epsscale{1.0} \plotone{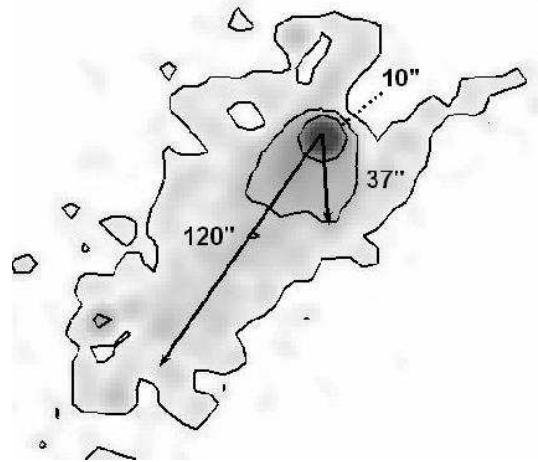}
\caption{The contours and spatial scales for the three main emission regions in the \textit{Chandra} image; the compact source region inside the innermost contour, cometary structure inside the middle contour, and the extended X-ray emission inside the outermost contour. \label{morphcont}}
\end{figure}

At the tip of the radio finger, the \textit{Chandra} image shows a compact source whose spatial profile is significantly more extended than
that of a point source (see Figure \ref{chandra}b). Figure \ref{morphcont} shows the contours and scales for the three main emission regions in the Chandra image, the compact source region in the innermost contour, the cometary region in the middle contour, and the elongated X-ray emission that is coincident with the radio finger in the outer contour. We used Sherpa to spatially fit the unbinned image of the compact source region
in the 0.3 to 9.0 keV energy band. The fitting was performed on a
circular section of the image, 24 pixels ($\sim$ 12 \arcsec) in radius, centered on the brightest region of the compact source. A normalized image of a 3 keV point
spread function, generated with the Ciao tool \textit{mkpsf}, was
used as a convolution kernel in the fitting. We used a model consisting of a delta function plus a Gaussian, and a level offset for the local background surrounding the compact source. The spatial profile and the best fit model are shown in Figure \ref{psplot}a. The best fit model consists of a delta function with an amplitude of 48$\pm$19 $counts$, a Gaussian with a fwhm of 5\farcs5$\pm$0\farcs5, and an ellipticity of 0.21$\pm$0.08, and a background offset of 1.1 $counts\:arcsec^{-2}$. The quoted unceratinties represent 1-sigma errors on the fitted parameters. The center coordinates of the delta function and the Gaussian components are 15:54:24.5, -55:03:45.1, and  15:54:24.4, -55:03:44.6, respectively. The total fluxes inside the 12 \arcsec radius aperture are 48 counts in the point source component and 749 $counts$ in the extended, Gaussian component. The best-fit values are summarized in Table \ref{compacttab}. Figure \ref{psimage} shows the \textit{Chandra} image of the compact source region in the left panel, the image of the best fit model in the right panel, and the residual image in the middle panel.  We note that the residual image shows a slightly elongated knot of emission, approximately 4\arcsec in length, that originates near the peak
of the point source and extends to the northwest. The residual
emission in the brightest pixels of the knot is approximately 5
sigma above the noise.


The results of the fitting suggest that the emission originates from
a point source, assumed to be the pulsar,  embedded in a more extended structure which accounts for 94 \% of the compact source flux . A similar type of emission was seen in the Mouse \citep{gae04} bow shock nebula, where the X-ray
emission from its compact source is also composed of a point source embedded in a broader component,
2\farcs4 in size. In the Mouse, this emission is contained within
the termination shock of the pulsar and it was attributed to X-ray knots
that are produced in its vicinity \citep{gae04}. In
the case of G327.1-1.1, the size of the extended component is much
broader, $\sim$12\arcsec in diameter, and the emission is surprisingly
uniform. A possible explanation for the origin of this
emission is the shocked pulsar wind downstream of the termination
shock, which would require the boundary of the shock to
be on a much smaller scale than the spatial extent of the compact source and
well within the innermost contour of Figure \ref{morphcont}.

\subsubsection{Prong and Bubble Structures}

The most unusual feature in the \textit{Chandra} and \textit{XMM}
images are two prong-like structures that extend into a faint
bubble, approximately 3 arcminutes in diameter (Figures \ref{chandra} and
\ref{bow}). The prongs are approximately 1.5 arcminutes in length
and their axes are not
aligned with the position of the compact source, which is presumably
the pulsar. It seems unlikely that the prongs and bubble structures are from
the SNR itself and that their position only coincidentally falls
along the line of sight to the position of the pulsar and is aligned
along the direction of the pulsar motion. 
The structures may arise from the violent interaction between the reverse shock and the PWN that would cause the pulsar wind material to mix in with the SNR ejecta.


\begin{figure*}
\epsscale{1.0} \plotone{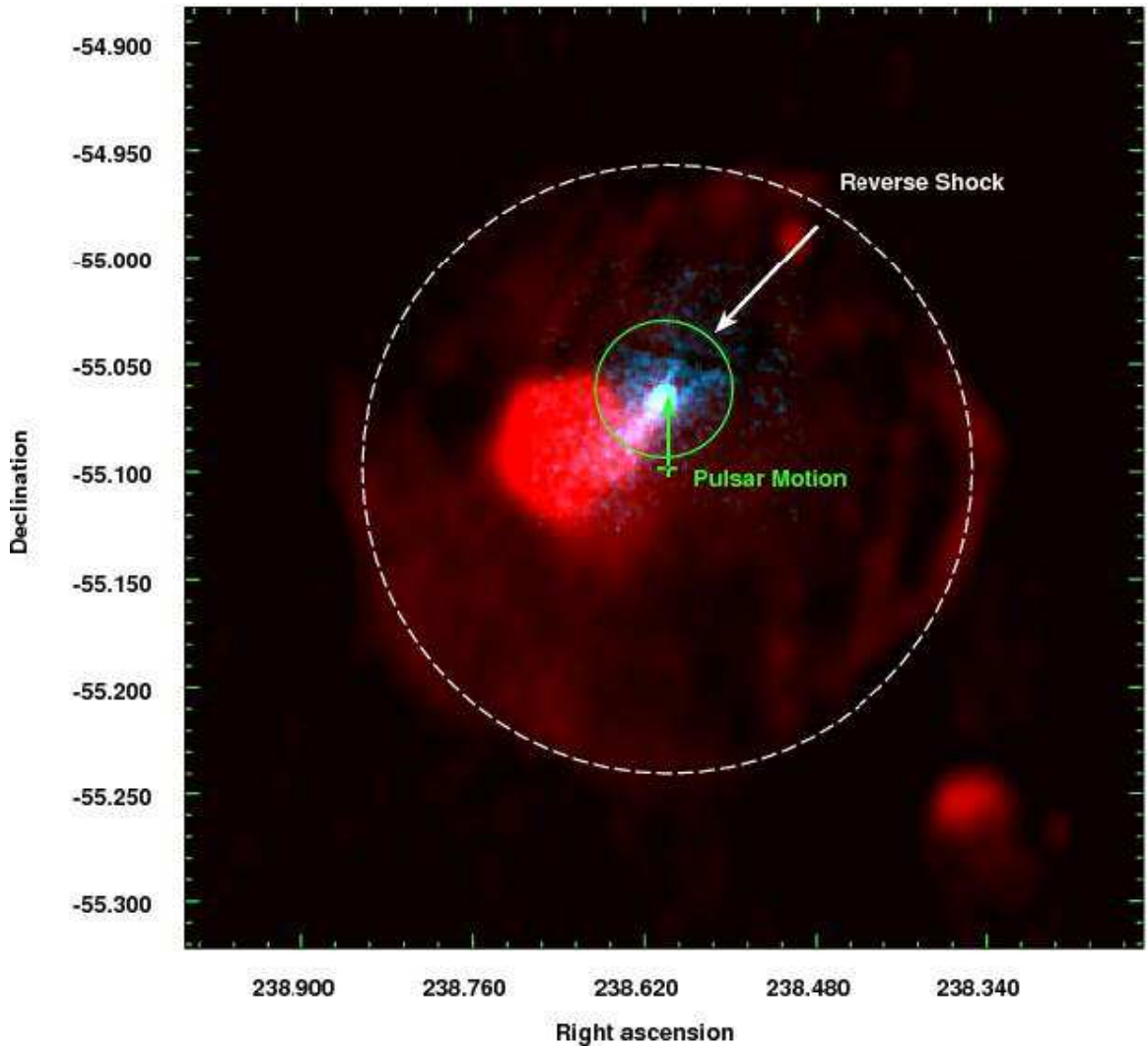}
\caption{\label{bubble} The MOST radio image is show in red and
Chandra ACIS X-ray image in blue/green. The figure illustrates a possible physical scenario that could give rise to the observed radio and X-ray morphology. The white dashed circle is the boundary of the radio shell, the green cross is the geometric center of the shell, the green arrow represents a possible direction of the pulsar motion, the green circle is the position of the PWN immediately before the passage of the reverse shock, and the white arrow indicates the proposed direction of the reverse shock.}
\end{figure*}


\subsubsection{Size of the PWN and Pulsar Wind Cavity}\label{pwn_size}

In this section, we estimate the exptected sizes of the PWN and the pulsar wind cavity for two scenarios and compare them to the observations; the case where the PWN has reformed and is now expanding subsonically, and the case where the PWN has been deformed into a bow shock, due to the pulsar motion.
In Section \ref{nonthermal} we found that the timescale for the
reverse shock to interact with the entire PWN surface of G327.1-1.1
is on the same order as the remnant age. We note that the uncertainty on these estimates is high and that it is possible that the PWN has entered the subsonic stage of expansion. Here we assume that the PWN is expanding
subsonically, and has reached a pressure equilibrium with the SNR, after the passage of the reverse shock, in order to make rough estimates of the relative sizes of the SNR radius,
(R$_{SNR}$), the PWN radius (R$_{PWN}$), and the termination shock radius (R$_{ts}$). Following \cite{swa01} and setting $P_{\rm pwn}$ equal to $P_{\rm snr}$, the radius of the PWN can be expressed as

\begin{equation}
{R}_{PWN}\simeq \left(\frac{\dot{E}t}{E_0}\right)^{1/3}{R}_{SNR},
\label{Rpwn}
\end{equation}

\noindent where $\dot{E}$ is the mechanical luminosity driving the pulsar wind, found in Section \ref{nonthermal},  and $t$ is the time passed since the passage of the reverse shock ($t_{SNR}$ minus $t_{col}$). In Section \ref{evolution}, we calculated that $t_{SNR}$ and $t_{col}$ are on the same order of magnitude, but the values are associated with very large uncertainties. For the purpose of our calculation, we set $t$ equal to 5000 yr. Setting the pressure of
the relativistic wind terminated at a radius $R_{\rm ts}$ \citep{ken84} equal to the SNR pressure, we obtain the following
expression for $R_{\rm ts}$

\begin{equation}
R_{ts}\simeq\sqrt{\frac{\dot{E}}{cE}}R_{snr}^{3/2}.
\end{equation}

Using the parameters for G327.1-1.1, and our estimate of $\dot{E}$ from Section \ref{nonthermal},
we estimate that $R_{PWN}\simeq2.7pc\:t^{1/3}_{1000\:yr}$, or $\sim60\arcsec\:t^{-1/3}_{1000\:yr}$. Since the average extent of the X-ray emission observed by \textit{Chandra} is on the order of 60\arcsec (see Figure \ref{morphcont}), the time since the reverse shock collided with the PWN should be $\sim1000\:yr$. The prongs and bubble structures ahead of the pulsar may be caused by the turbulant interaction between the reverse shock and the PWN, and may provide further evidence for a somewhat recent reverse shock collision. We calculate a termination shock
radius of $0.25\:pc$, or 6\arcsec, roughly the same size as the extent of the compact source seen in the \textit{Chandra} image (indicated by the innermost contour of Figure \ref{morphcont}). Even though the estimated values are only
first order estimates, the predicted size of $R_{ts}$ indicates that the pulsar wind cavity is too small to be associated with the observed cometary structure. In this scenario, the cometary morphology may be caused by the passage of the reverse shock, which disrupted the reformed PWN from the NW direction.

As an alternative, we consider the case in which the PWN has been deformed into a bow shock and make similar estimates to compare with our observations. We assume a pulsar velocity of 770 km/s, the minimum required velocity for bow shock formation, found in Section \ref{pulsar_vel}. Following the discussion by \citet{swa03}, we calculate the values of the pressure behind the SNR blast wave, $P_{sh}$, pulsar mach number, $M_{psr}$, bow shock pressure, $P_{bs}$, leading to the the radius of the forward termination shock, $R_{ts}^F$ (equations 6, 10, 16, and 21 of \cite{swa03}). Using the parameters for G327.1-1.1 and $\gamma$ of 5/3, we calculate that $P_{sh}=5.2\times10^{-10}dyne\:cm^{-2}$, $M_{psr}=2.8$, $P_{bs}=4.7\times10^{-9}dyne\:cm^{-2}$, and finally $R_{ts}^F=0.033\:pc$, or 0\farcs8, based on a distance of 9.0 kpc. The radius of the backward termination shock would then be on the order of 2\farcs7, making the entire extent of the termination shock $\sim$3\farcs5 \citep{buc02,swa03}. This is somewhat smaller, but on the same order of magnitude as the size of the compact source in G327.1-1.1. The expected size of the contact discontinuity radius is $\approx1.33R_{ts}^F=0.04\:pc$, or 1\arcsec \citep{buc02,swa03,gae04}. Based on these estimates, the cometary structure in the \textit{Chandra} image appears too large to be attributed to the contact discontinuity, since it extends at least 8\arcsec beyond the extent of the compact source in the forward direction (see Figure \ref{morphcont}). The relative sizes of the observed structures are not consistent with estimated sizes from bow shock simulations, but due to the large uncertainties in the derived parameters for G327.1-1.1, we cannot rule out the bow shock scenario on this basis alone. However, we note that the observed X-ray emission ahead of the pulsar in the form of prongs and bubble structures, does suggest that the cometary structure that we are observing is most likely not a bow shock.

\subsubsection{Possible Physical Scenario}

There are two possible scenarios that may give rise to the
cometary morphology observed in G327.1-1.1; the newly forming
PWN may be deformed into a bow shock when the pulsar velocity
becomes supersonic, or an asymmetric passage of the reverse shock
may disrupt the PWN from one side and give rise to the cometary
morphology. While we were not able to rule out either scenario, the
latter appears to be a more likely based on the observed X-ray emission ahead of the pular (see Section \ref{pwn_size}). We
suggest that PWN in G327.1-1.1 is in the subsonic stage of
expansion, after the passage of the reverse shock. Figure \ref{bubble}
illustrates a possible physical scenario that could explain the
observed morphology. If we assume that the pulsar was born in the
geometric center of the radio shell, indicated by a white dashed circle, 
its current position suggests that the velocity vector is in
the north direction, as indicated by the green arrow. Before the
passage of the reverse shock, the PWN was expanding supersonically
and was being carried along with the moving pulsar to the north. The
position of the PWN immediately before the passage of the reverse
shock would therefore be north of the remnant center, as shown by the green
circle in Figure \ref{bubble}. If inhomogeneities in the ISM caused the
reverse shock to reach the PWN from the NW direction first, it would
displace the PWN to the location of the radio relic and cause the
newly forming PWN to take on the observed cometary
morphology \citep{swa04}. Hydrodynamic simulations of \citet{swa04}
show the various stages of this process, with the main difference
being that an additional asymmetry, besides the pulsar's velocity
vector, is required to explain the morphology. In order for the reverse shock to first reach the PWN surface from the northwest direction, a higher ISM density in the West would be required, which cannot be confirmed or ruled out by our data.

\section{CONCLUSIONS} \label{concl}

In this paper, we analyzed the \textit{Chandra} and \textit{XMM}
imaging and spectroscopy of G327.1-1.1 in order to characterize the
nature of the X-ray emission, determine the properties of the SNR
and the pulsar progenitor, and understand the evolutionary and
physical scenario that would lead to the observed morphology.
The X-ray images show an extended compact source that is embedded in a
cometary structure, from which a trail of emission extends towards the bright radio PWN.
Prong-like structures originate near the compact source and
extend into a faint bubble, 3 arcminutes in diameter. A symmetric
shell coincident with the the shell observed in the radio is evident
in the 1-2 keV \textit{XMM} image.
The X-ray spectrum of the PWN is described
by a power-law model with an average photon index of 2.11 $\pm$ 0.03. The shell emission is best described by thermal
model with a temperature of 0.30 $\pm$ 0.01 keV. Using the Sedov model and assuming a distance of 9 kpc, we calculate a remnant radius of 22 pc, an age of 18000 years, shock velocity of
500 $km/s$, n$_0$ of 0.12 $cm^{-3}$, a swept-up mass of 31
$M_{sol}$, and an explosion energy of 0.5 x 10$^{51}$ ergs. Since
the remnant has most likely expanded into a nonuniform medium, the
derived dynamical properties are expected to vary.

Based on the timescale for reverse shock interaction and the derived
remnant age, it is possible that G327.1-1.1 is in the subsonic
expansion stage of its evolution. While we can not discount the
possibility of a bow shock formation in G327.1-1.1 based on the 
calculations of the relative sizes of the PWN and the termination
shock radii, the observed X-ray emission ahead of the pulsar makes the bow shock scenario less likely. The cometary morphology may be explained by the passage of the reverse shock from the northwest
direction that disrupted the newly forming PWN, which was originally
located north of the SNR center.

\acknowledgments

Support for this work is provided by NASA Grant GO6-7053X. B.M.G. acknowledges support from the Australian Research Council through a Federation Fellowship (grant FF0561298). P.O.S. acknowledges support from NASA contract NAS8-03060.

\bibliographystyle{plain}
















\end{document}